\documentclass[aps,preprint]{revtex4}%
\usepackage{amsfonts}
\usepackage{amsmath}
\usepackage{amssymb}
\usepackage{graphicx}%
\begin{document}
\preprint{ }
\title[spin squeezing]{Spin squeezing via atom - cavity field coupling}
\author{Claudiu Genes and P. R. Berman}
\affiliation{Michigan Center for Theoretical Physics, FOCUS Center, and Physics Department,
University of Michigan, Ann Arbor, Michigan \ 48109-1120}
\author{A. G. Rojo}
\affiliation{Department of Physics,Oakland University, Rochester, Michigan 48309}
\author{}

\begin{abstract}
Spin squeezing via atom-field interactions is considered within the context of
the Tavis-Cummings model. An ensemble of $N$ two-level atoms interacts with a
quantized cavity field. For all the atoms initially in their ground states, it
is shown that spin squeezing of both the atoms and the field can be achieved
provided the initial state of the cavity field has coherence between number
states differing by 2. Most of the discussion is restricted to the case of a
cavity field initially in a coherent state, but initial squeezed states for
the field are also discussed. Optimal conditions for obtaining squeezing are
obtained. An analytic solution is found that is valid in the limit that the
number of atoms is much greater than unity and is also much larger than the
average number of photons, $\alpha^{2}$, inititally in the coherent state of
the cavity field. In this limit, the degree of spin squeezing increases with
increasing $\alpha$, even though the field more closely resembles a classical
field for which no spin squeezing could be achieved.

\end{abstract}
\volumeyear{year}
\volumenumber{number}
\issuenumber{number}
\eid{identifier}
\date[Date text]{date}
\received[Received text]{date}

\revised[Revised text]{date}

\accepted[Accepted text]{date}

\published[Published text]{date}

\startpage{1}
\endpage{ }
\maketitle

\section{\textbf{Introduction}}

Spin squeezed states offer an interesting possibility for reducing quantum
noise in precision measurements \cite{wineland,theory,experiment}. Spin
squeezing is described in terms of spin operators that are associated with
quantum mechanical operators of two-level atoms (TLA) (we refer to atoms and
spins interchangeably). In an appropriate interaction representation,
combinations of atomic raising and lowering operators for atom $j$ are
associated with the $x$ and $y$ spin components ($S_{x}^{j}$ and $S_{y}^{j})$,
while the population difference operator for the two states is associated with
the $z$ spin component ($S_{z}^{j})$. One then defines collective operators
$S_{\alpha}=\sum_{j}S_{\alpha}^{j}$ that obey the usual spin commutator
relations. If one measures an average spin $|\left\langle \mathbf{S}%
\right\rangle |=\sqrt{\left\langle S_{x}\right\rangle ^{2}+\left\langle
S_{y}\right\rangle ^{2}+\left\langle S_{z}\right\rangle ^{2}}$ then the system
is said to be spin-squeezed if%
\begin{equation}
\xi_{\perp}=\sqrt{2S}\Delta S_{\perp}/|\left\langle \mathbf{S}\right\rangle
|<1, \label{sqzc}%
\end{equation}
where $\Delta S_{\perp}$ is the uncertainty in a spin component perpendicular
to $\left\langle \mathbf{S}\right\rangle $, $S=N/2,$ and $N$ is the number of
atoms \cite{wineland,theory}. Spin squeezing is impossible for a single atom
and requires the entanglement of the spins of two or more atoms. There are
many ways to theoretically construct a Hamiltonian that can give rise to the
necessary entanglement among N two-level atoms. Since a linear Hamiltonian
merely rotates the spin components leaving the uncertainties unchanged, it is
generally necessary to use Hamiltonians that are quadratic in the spin
operators to generate squeezing. On the other hand, it is possible to generate
squeezing using a Hamiltonian linear in the spin operators provided the spin
system is coupled to another quantum system, such as a harmonic oscillator. It
is then not surprising to find that a squeezed state of the oscillator can be
transferred to some degree to the atoms. What may be a little more surprising
is that an oscillator prepared in a coherent state and coupled to the spins
can result in spin squeezing. In this paper, we study the dynamics of the
creation of squeezing in an ensemble of spins via coupling to a cavity field
in the Tavis-Cummings model \cite{tavis}. An ensemble of $N$ atoms is coupled
in a spatially independent manner to the $N$ atoms with no losses for the
field and with the neglect of any spontaneous emission for the atoms. We are
concerned mainly with the type of spin squeezing that can be generated by
coupling to a radiation field that is initially a coherent state, but also
will consider an initial state of the field that is a squeezed state. The
evolution of the radiation field will also be determined. There have been a
number of studies of atom-field dynamics in the Tavis-Cummings model in which
the squeezing of the cavity field was calculated in various limits
\cite{prevtc}. Some numerical solutions to the problem of spin squeezing in
the Tavis-Cummings model are given in Ref. \cite{wineland}.

The initial condition for the atoms is taken as one in which all the atoms are
in their lower energy state, corresponding to a coherent spin state. For a
very large number of atoms ($N\gg1$ and $N$ much greater than the average
number of photons in the coherent state of the field), the relevant energy
levels of the spin system approach those of a simple harmonic oscillator with
corrections that vanish as $N\sim\infty$. Thus it would seem that spin
squeezing can never be achieved if the initial state of the cavity field is a
coherent state, since one is dealing with a linear interaction between two
harmonic oscillators each of them initially in a coherent state. Nevertheless,
we show that for any finite $N,$ spin squeezing occurs and the degree of spin
squeezing actually increases with increasing field strength.

To follow the atom-field dynamics, we consider first a system having $N=2$. It
is not difficult to obtain analytic solutions in this case, enabling us to
track the dependence of $\xi_{\perp}$ on field strength and $N$. In addition,
we determine if the squeezed vacuum state results in optimal transfer of
squeezing from the fields to the atoms. After discussing the two atom case, we
generalize the results to $N$ atoms.

The paper is organized as follows. In Sec. II we present the mathematical
framework and obtain results that show that no squeezing can be achieved when
the field is either classical, or quantized in a number state. In Sec. III, we
consider the $N=2$ case and obtain analytical results for both coherent and
squeezed cavity fields, in the limit that the average number of photons in the
field is much less than unity. Numerical solutions for larger field strength
are presented. In Sec. IV, the results are generalized to $N$ atoms. In both
sections III and IV, the time evolution and squeezing of the field is also
calculated for the case that the field is initially in a coherent state. In
Sec. V, a formal derivation of the large $N$ limit is given using the
Holstein-Primakoff transformation \cite{holstein}, valid for an arbitrary
strength of the coherent cavity field. The Holstein-Primakoff \ transformation
was used previously by Persico and Vetri \cite{persico} to analyze the
atom-field dynamics in the limit of large $N$. The approach we follow differs
somewhat from theirs and our results seem to have a wider range of validity
than that stated by Persico and Vetri. The results are summarized in Sec. VI.

\section{\textbf{\ General Considerations}}

In dipole and rotating-wave approximations, the Hamiltonian for an ensemble of
TLA (lower state $\left\vert 1\right\rangle $, upper state $\left\vert
2\right\rangle ,$ transition frequency $\omega$) interacting with a resonant
cavity field, $E(t)=Ea\,e^{-i\omega t}+E^{\ast}a\,^{\dag}e^{i\omega t}$, is of
the form%

\[
H=\hbar\omega S_{z}+\hbar\omega a^{+}a+\hbar g(S_{+}a+S_{-}a^{+}),
\]
where $S_{z}=\sum_{j=1}^{N}\left[  \left(  \left\vert 2\right\rangle
\left\langle 2\right\vert \right)  _{j}-\left(  \left\vert 1\right\rangle
\left\langle 1\right\vert \right)  _{j}\right]  /2$, $S_{+}=\sum_{j=1}%
^{N}\left(  \left\vert 2\right\rangle \left\langle 1\right\vert \right)
_{j}\,e^{-i\omega t},$ $S_{-}=\sum_{j=1}^{N}\,\left(  \left\vert
1\right\rangle \left\langle 2\right\vert \right)  _{j}e^{i\omega t},$
$S_{x}=\left(  S_{+}+S_{-}\right)  /2$, $S_{y}=\left(  S_{+}-S_{-}\right)
/2i$, $a$ and $a^{\dagger}$ are annihilation and creation operators for the
field, and $g$ is a coupling constant. The spin operators have been defined in
a reference frame rotating at the field frequency. Constants of the motion are
$S^{2}=$ $S_{x}^{2}+S_{y}^{2}+S_{z}^{2}$ and $\left(  S_{z}+a^{+}a\right)  $.
If, initially, all spins are in their lower energy state, then $S^{2}%
=N^{2}/4.$ In order to calculate $\xi_{\perp}$ from Eq. (\ref{sqzc}), one must
first find $\left\langle \mathbf{S}\right\rangle $ and define two independent
directions orthogonal to $\left\langle \mathbf{S}\right\rangle ,$ $S_{\perp1}$
and $S_{\perp2}$. It then follows that $\left\langle S_{\perp1}\right\rangle
=\left\langle S_{\perp2}\right\rangle =0$ and
\[
\left(  \Delta S_{\perp i}\right)  ^{2}=\frac{N}{4}+\sum_{j,j^{\prime}\neq
j}\left\langle S_{\perp i}^{(j)}S_{\perp i}^{(j^{\prime})}\right\rangle ,
\]
where $i=1,2$ and $S^{(j)}$ is a spin operator for atom $j.$

A necessary condition to have $\xi_{\perp}<1$ is that the different spins are
entangled. To see this, take a system in which $\left\langle \mathbf{S}%
\right\rangle $ is aligned along the $z$ axis, with the $x$ axis is chosen
such that $\Delta S_{x}$ is the minimum value of $S_{\perp}$. Using the facts
that $\left\langle \mathbf{S}\right\rangle =S_{z}$, $\left\langle
S_{x}\right\rangle =\left\langle S_{y}\right\rangle =0,$ $\Delta S_{x}\Delta
S_{y}\geq\left\vert \left\langle S_{z}\right\rangle \right\vert /2$, one finds%
\[
\xi_{x}=\sqrt{N}\Delta S_{x}/|\left\langle S_{z}\right\rangle |\geq\sqrt
{N}/\Delta S_{y}=\left[  1+\sum_{j,j^{\prime}\neq j}\left\langle S_{y}%
^{(j)}S_{y}^{(j^{\prime})}\right\rangle \right]  ^{-1}%
\]
For correlated states, the sum can be positive and one cannot rule out the
possibility that $\xi_{x}<1.$ On the other hand, for uncorrelated states,
using the fact that $\left\langle S_{y}\right\rangle ^{2}=0$, it follows that
$1+\sum_{j,j^{\prime}\neq j}\left\langle S_{y}^{(j)}S_{y}^{(j^{\prime}%
)}\right\rangle =1-\sum_{j}\left\langle S_{y}^{(j)}\right\rangle ^{2}.$ As a
consequence, $\xi_{x}\geq1$ and there is no spin squeezing for uncorrelated states.

We note two general conclusions that are valid for arbitrary $N.$ First, if we
were to replace the cavity field by a classical field, the Hamiltonian would
be transformed into
\[
H_{class}=\sum_{j}\left[  \hbar\omega S_{z}^{(j)}+\hbar g^{\prime}(S_{+}%
^{(j)}e^{-i\omega t}+S_{-}^{(j)}e^{i\omega t})\right]  ,
\]
where $g^{\prime}$ is a constant. Since the Hamiltonian is now a sum of
Hamiltonians for the individual atoms, the wave function is a direct product
of the wave functions of the individual atoms. As a consequence, there is no
entanglement and no spin squeezing for a classical field. Second, if the
initial state of the field is a Fock state, although there is entanglement
between the atoms and the field, there is no spin squeezing. There is no spin
squeezing unless the initial state of the field has coherence between at least
two states differing in $n$ by 2.\textbf{\ }For a Fock state, there is no such
coherence and $\xi_{\perp}\geq1$.

It is convenient to carry out the calculations in an interaction
representation with the wave function expressed as%
\begin{equation}
\left\vert \psi(t)\right\rangle =\overset{N/2}{\underset{m=-N/2}{\sum}%
}\overset{\infty}{\underset{n=0}{\quad\sum}}c_{mk}(t)\,e^{-i\omega\left(
m+n\right)  t}\left\vert m,n\right\rangle , \label{wf}%
\end{equation}
where $m$ labels the value of $S_{z}$ and $n$ labels the number of photons in
the cavity field. In this representation, the Hamiltonian governing the time
evolution of the $c_{mk}(t)$ is given by%
\begin{equation}
H=\hbar g(S_{+}a+S_{-}a^{+}). \label{hamil}%
\end{equation}

\section{N=2}

We first set $N=2,$ $S=1$. If the spins are all in their lower energy state at
$t=0$, the initial wave function is
\begin{equation}
\left\vert \psi(0)\right\rangle =\sum_{k=0}^{\infty}\overset{}{c_{k}%
}\left\vert -1,k\right\rangle , \label{init}%
\end{equation}
where the $c_{k}$ are the initial state amplitudes for the field. Solving the
time-dependent Schr\"{o}dinger equation with initial condition (\ref{init}),
one finds%

\begin{subequations}
\label{cev}%
\begin{align}
c_{-1,k}(t)  &  =\frac{1}{(2k-1)}\left[  k-1+k\cos(\sqrt{4k-2}gt)\right]
c_{k}\label{cevc}\\
c_{0,k}(t)  &  =-i\sqrt{\frac{k+1}{2k+1}}\sin(\sqrt{4k+2}gt)c_{k+1}%
\label{cevd}\\
c_{1,k}(t)  &  =\frac{\sqrt{\left(  k+1\right)  (k+1)}}{2k+3}\left[
-1+\cos(\sqrt{4k+6}gt)\right]  c_{k+2}. \label{ceve}%
\end{align}
These state amplitudes can be used to calculate all expectation values of the
spin operators.

\subsection{\textbf{Coherent State}}

If the initial state of the cavity field is a coherent state, then
\end{subequations}
\begin{equation}
c_{k}=\alpha^{k}e^{-\left\vert \alpha\right\vert ^{2}/2}/\sqrt{k!},
\label{coh}%
\end{equation}
and the average number, $n_{0}$, of photons in the field is given by $n_{0}=$
$\left\vert \alpha\right\vert ^{2}.$ For simplicity, we take $\alpha$ and $g$
to be real.

\subsubsection{\textbf{\bigskip Solution for }$\left\vert \alpha\right\vert
^{2}<<1$}

\begin{figure}
[ptb]
\begin{center}
\label{Fig. 1}
\includegraphics[
height=2.6135in,
width=4.0456in
]%
{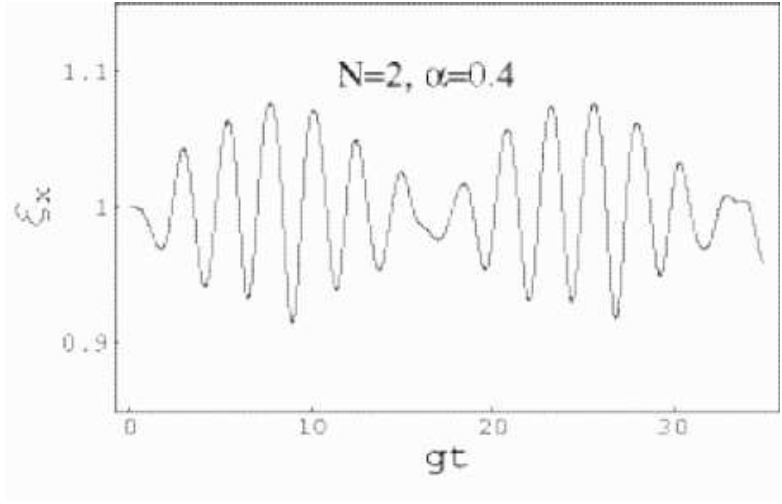}%
\caption{Spin squeezing $\xi_{x}$ as a function of $gt$ for $\alpha=0.4$ and
$N=2$.}%
\end{center}
\end{figure}

Keeping terms to order $\alpha^{2},$ one finds from Eqs. (\ref{cev}) and
(\ref{coh}) that the only state amplitudes of importance are%

\begin{subequations}
\label{cn}%
\begin{align}
c_{-1,0}(t)  &  =(1-\alpha^{2}/2)\label{cna}\\
c_{-1,1}(t)  &  =\alpha\cos(\sqrt{2}gt)\label{cnb}\\
\ c_{-1,2}(t)  &  =\frac{\alpha^{2}}{3\sqrt{2}}\left[  1+2\cos(\sqrt
{6}gt)\right] \label{cnc}\\
c_{0,0}(t)  &  =-i\alpha\sin(\sqrt{2}gt)\label{cnd}\\
c_{0,1}(t)  &  =-\frac{i\alpha^{2}}{\sqrt{3}}\sin(\sqrt{6}gt)\label{cne}\\
c_{1,0}(t)  &  =-\frac{\alpha^{2}}{3}\left[  1-\cos(\sqrt{6}gt)\right]  .
\label{cnf}%
\end{align}
The spin components' expectation values are:
\end{subequations}
\begin{subequations}
\label{spin}%
\begin{align}
\left\langle S_{x}\right\rangle  &  =0\qquad\quad\left\langle S_{y}%
\right\rangle =\sqrt{2}\alpha\sin(\sqrt{2}gt)\ \label{spin1}\\
\left\langle S_{z}\right\rangle  &  =-\left[  1-\alpha^{2}\sin^{2}(\sqrt
{2}gt)\right]  \ . \label{spin2}%
\end{align}

The motion of the average value for the spin vector operator is in the $yz$
plane, with the length of the vector always equal to unity, to order
$\alpha^{2}$. Since $\left\langle S_{x}\right\rangle =0$, the plane in which
we look for spin squeezing is the one defined by the $x$ axis and an axis
orthogonal to both $\mathbf{\hat{x}}$ and the instantaneous direction of the
spin. Making the appropriate rotation in the $yz$ plane to define a
$y^{\prime}$ axes perpendicular to $\left\langle \mathbf{S}\right\rangle $ and
$\mathbf{\hat{x}},$ and afterwards choosing an arbitrary direction defined by
an angle $\phi$ in this plane, one finds that $\xi_{\phi}\geqslant\min\{$
$\xi_{x},$ $\xi_{y^{\prime}}\}$, which implies that the best squeezing is to
be found in either the $x$ or $y^{\prime}$ directions. The analytical
expressions for $\xi_{x},$ $\xi_{y^{\prime}}$ are:%

\end{subequations}
\begin{subequations}
\label{squeez}%
\begin{align}
\xi_{x}  &  =\sqrt{2}\frac{\Delta S_{x}}{|\left\langle \mathbf{S}\right\rangle
|}\simeq1+\alpha^{2}\left\{  \frac{1}{2}\sin^{2}(\sqrt{2}gt)-\frac{2}{3}%
\sin^{2}\left(  \sqrt{6}gt/2\right)  \right\} \label{squeez1}\\
\xi_{y^{\prime}}  &  =\sqrt{2}\frac{\Delta S_{y^{\prime}}}{|\left\langle
\mathbf{S}\right\rangle |}\simeq1+\alpha^{2}\left\{  -\frac{1}{2}\sin
^{2}(\sqrt{2}gt)+\frac{2}{3}\sin^{2}\left(  \sqrt{6}gt/2\right)  \right\}
\label{squeez2}%
\end{align}
The lowest possible value for the squeezing occurs in the $x$ direction and is
equal to%

\end{subequations}
\begin{equation}
\xi_{\min}=1-\frac{2}{3}\alpha^{2} \label{smin}%
\end{equation}
at a time when $\sin(\sqrt{2}gt)=0$ and $\cos(\sqrt{6}gt)=-1$. The squeezing
$\xi_{x}$ as a function of $gt$ for $\alpha=0.4$ is plotted in Fig.
\ref{Fig. 1}.%

\subsubsection{\bigskip\textbf{Numerical results for all values of }$\alpha$}

General expressions for the spin expectation values and variances can be
obtained and used for numerical simulations for any values of $\alpha$. With
$\alpha$ real, the expectation value of the $x$ component of the spin vanishes
and, with the notation $\overline{c_{0,n}}=\frac{c_{0,n}}{i},$%

\begin{align*}
\left\langle S_{y}\right\rangle  &  =\sqrt{2}\overset{\infty}{\underset
{n=0}{\sum}}\overline{c_{0,n}}(c_{1,n}-c_{-1,n})\\
\left\langle S_{z}\right\rangle  &  =\overset{\infty}{\underset{n=0}{\sum}%
}(\left\vert c_{1,n}\right\vert ^{2}-\left\vert c_{-1,n}\right\vert ^{2})
\end{align*}
The variances are:%

\begin{subequations}
\label{sx}%
\begin{align}
\left(  \Delta S_{x}\right)  ^{2}  &  =\left\langle S_{x}^{2}\right\rangle
=\frac{1}{2}+\overset{\infty}{\underset{n=0}{\sum}}\{\frac{1}{2}\left\vert
c_{0,n}\right\vert ^{2}+c_{1,n}c_{-1,n}\}\label{sxa}\\
\left(  \Delta S_{y}\right)  ^{2}  &  =\left\langle S_{y}^{2}\right\rangle
-\left\langle S_{y}\right\rangle ^{2}=\frac{1}{2}+\underset{n=0}%
{\overset{\infty}{\sum}}\{\frac{1}{2}\left\vert c_{0,n}\right\vert
^{2}-c_{1,n}c_{-1,n}\}-\left\langle S_{y}\right\rangle ^{2} \label{sxb}%
\end{align}
The variance in the $x$ component of the spin cannot be less than 1/2 unless
$c_{1,n}c_{-1,n}<0$. Since $c_{1,n}c_{-1,n}$ is proportional to $c_{k+2}%
c_{k},$ where the $c_{k}$s are initial state amplitudes for the cavity field,
spin squeezing can be induced by a field only if the field has at least one
nonvanishing off-diagonal density matrix element $\rho_{kk^{\prime}}$ for
which $\left\vert k-k^{\prime}\right\vert =2.$

The values for the spin averages and uncertainties are calculated in terms of
$\alpha$ and $gt$. For $\alpha^{2}<<1$ the numerical and analytical results
agree. For larger values of $\alpha,$ no analytical solution is available. The
numerical results indicate that the optimal squeezing is obtained in the
$\mathbf{\hat{x}}$ direction. As $\alpha$ is increased, the spin squeezing
increases and then decreases for $\alpha\gtrsim0.9,$ as shown in Fig.
\ref{Fig. 2}.%

\begin{figure}
[ptb]
\begin{center}
\label{Fig. 2}
\includegraphics[
trim=0.279353in 0.356854in 0.000000in 0.000000in,
natheight=3.338200in,
natwidth=3.940100in,
height=3.0234in,
width=3.7075in
]%
{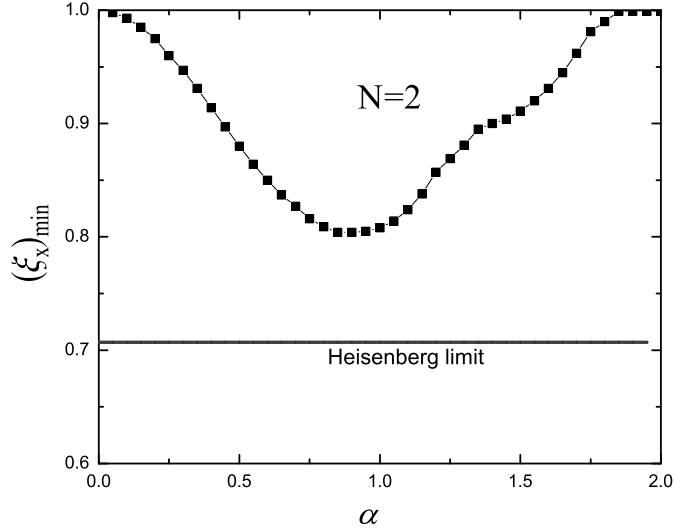}%
\caption{Optimal spin squeezing $\left(  \xi_{x}\right)  _{\min}$ as a
function of $\alpha$ for $N=2$. The time range out to $gt=5000$ was explored
in obtaining the minimal squeezing. In this and other plots, the point
represent actual values for which the squeezing was calculated. A line is
drawn through these points.}%
\end{center}
\end{figure}
With increasing $\alpha,$ the optimal squeezing occurs at increasingly large
values of $gt$. For example, with $\alpha=1.6,$ there is effectively no spin
squeezing for $gt<333$ and the optimal spin squeezing occurs for $gt=2439$.
The squeezing data in this and subsequent graphs is the optimal squeezing that
is obtained for $gt\ $less than some arbitrary cutoff that we have chosen. In
the limit of $\alpha\gg1$, the field closely resembles a classical field and
$\left(  \xi_{x}\right)  _{\min}$approaches unity. Formally, this result could
be derived by using a transformation proposed by Mollow \cite{mollow} in which
the transformed Hamiltonian is that of a classical field having amplitude
$\alpha$ plus a fluctuating field. Any spin squeezing that is produced depends
on the ratio of the fluctuations to the average field strength and must
decrease with increasing $\alpha$, provided the average number of photons in
the field is much larger than $N.$

\subsection{\textbf{Squeezing in the radiation field}}

Although the field is initially in a coherent state, it is squeezed as a
result of its interaction with the atoms \cite{prevtc}. In terms of quadrature
operators $\widehat{P}$ and $\widehat{Q}$ defined as:
\end{subequations}
\[
\widehat{Q}=\frac{1}{\sqrt{2}}(a+a^{+});\text{ \ \ }\widehat{P}=-\frac
{i}{\sqrt{2}}(a-a^{+})\text{\ }%
\]
with $[\widehat{Q},\widehat{P}]=i$, squeezing of the field occurs if the
variance of one of these two operators is smaller than the value it would have
for the vacuum field. Initially the field is in a coherent state of real
amplitude $\alpha$ with $\left\langle \widehat{Q}\right\rangle =\sqrt{2}%
\alpha$ and $\left\langle \widehat{P}\right\rangle =0,$ and variances
$(\Delta\widehat{Q})^{2}=(\Delta\widehat{P})^{2}=\frac{1}{2}$ satisfying the
minimum uncertainty condition
\begin{equation}
(\Delta\widehat{Q})^{2}(\Delta\widehat{P})^{2}=\frac{1}{4}\left\vert
\left\langle [\widehat{Q},\widehat{P}]\right\rangle \right\vert ^{2}=\frac
{1}{4}. \tag{21}%
\end{equation}
Using the wave function (\ref{wf}), one finds
\begin{align*}
\left\langle \widehat{Q}\right\rangle  &  =%
{\displaystyle\sum\limits_{m=-1}^{1}}
\sum_{k=0}^{\infty}\left[  \sqrt{k}c_{m,k-1}^{\ast}(t)+\sqrt{k+1}%
c_{m,k+1}^{\ast}(t)\right]  c_{m,k}(t)/\sqrt{2};\\
\left\langle \widehat{P}\right\rangle  &  =%
{\displaystyle\sum\limits_{m=-1}^{1}}
\sum_{k=0}^{\infty}\left[  \sqrt{k}c_{m,k-1}^{\ast}(t)+\sqrt{k+1}%
c_{m,k+1}^{\ast}(t)\right]  c_{m,k}(t)/\sqrt{2};\\
\left\langle \widehat{Q}^{2}\right\rangle  &  =\frac{1}{2}+\frac{1}{2}\left[
\sqrt{k(k-1)}c_{m,k-2}^{\ast}(t)c_{m,k}(t)+\sqrt{(k+1)(k+1)}c_{m,k+2}^{\ast
}(t)c_{m,k}(t)+2k\sqrt{k+1}\left\vert c_{m,k}(t)\right\vert ^{2}\right]  ;\\
\left\langle \widehat{P}^{2}\right\rangle  &  =\frac{1}{2}-\frac{1}{2}\left[
\sqrt{k(k-1)}c_{m,k-2}^{\ast}(t)c_{m,k}(t)+\sqrt{(k+1)(k+1)}c_{m,k+2}^{\ast
}(t)c_{m,k}(t)-2k\sqrt{k+1}\left\vert c_{m,k}(t)\right\vert ^{2}\right]  .
\end{align*}

To order $\alpha^{2}$, for the field initially in a coherent state, one finds
squeezing parameters%

\begin{align*}
\xi_{Q}  &  =\sqrt{2}\Delta\widehat{Q}\simeq1-\alpha^{2}\left\{  \cos
^{2}(\sqrt{2}gt)-\frac{1}{3}\left[  1+2\cos(\sqrt{6}gt)\right]  \right\} \\
\xi_{P}  &  =\sqrt{2}\Delta\widehat{P}\simeq1+\alpha^{2}\left\{  \cos
^{2}(\sqrt{2}gt)-\frac{1}{3}\left[  1+2\cos(\sqrt{6}gt)\right]  \right\}
\end{align*}
With this definition, squeezing occurs for $\xi_{Q}<1$ or $\xi_{P}<1$. To
second order in $\alpha$ the state of the field evolves in time as a minimum
uncertainty state but with squeezing transfer between the two quadratures. The
minimum value for the squeezing parameters that can be obtained is:%

\begin{subequations}
\label{fs}%
\begin{align}
(\xi_{Q})_{\min}  &  =1-\frac{4}{3}\alpha^{2}\label{fsa}\\
(\xi_{P})_{\min}  &  =1-\alpha^{2} \label{fsb}%
\end{align}
A continuous transfer of squeezing between the $Q$ quadrature and the $x$
component of the spin, and also between the $P$ quadrature and the $y$
component of the spin is taking place. The maximum field squeezing as a
function of $\alpha$ is shown in Fig. \ref{Fig. 3}.%

\begin{figure}
\label{Fig. 3}
[ptb]
\begin{center}
\includegraphics[
trim=0.248783in 0.199446in 0.000000in 0.000000in,
natheight=3.087400in,
natwidth=3.857100in,
height=2.9317in,
width=3.6538in
]%
{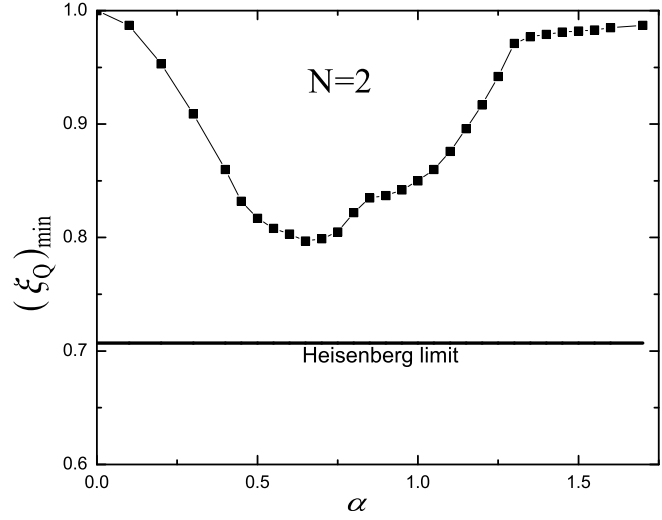}%
\caption{Optimal field squeezing $\left(  \xi_{Q}\right)  _{\min}$ as a
function of $\alpha$ for $N=2$. The time range out to $gt=5000$ was explored
in obtaining the minimal squeezing.}%
\end{center}
\end{figure}

\subsection{Squeezed initial cavity field}

>From Eq. (\ref{sx}) one can see that initial state coherence between photon
field states differing by 2 is needed for squeezing. The squeezed vacuum is a
superposition of even Fock states; therefore, it is a good choice for inducing
the necessary coherences in the atomic system. Analytical results are
available for a small squeezing parameter of the field, and numerical results
can be obtained for larger values. For a squeezed vacuum the $c_{k}$s are
given by%
\end{subequations}
\[
c_{0}=\frac{1}{\sqrt{\cosh r}};\text{ \ \ }c_{k}=\frac{\left(  k-1\right)
!!(-1)^{k/2}\tanh^{k/2}r}{\sqrt{k!\cosh r}}\text{ \ for }k\text{ even;
\ \ }c_{k}=0\text{ \ for }k\text{ odd,}%
\]
where $r$ is the squeezing parameter, assumed real. For any field containing
only even expansion coefficients, $\left\langle S_{x}\right\rangle
=\left\langle S_{x}\right\rangle =0.$ For $r\ll1$, one obtains for the spin squeezing%

\begin{align*}
\xi_{x}  &  =\sqrt{2}\frac{\Delta S_{x}}{|\left\langle \mathbf{S}\right\rangle
|}\simeq1+\frac{4}{3}r\sin^{2}(\sqrt{\frac{3}{2}}gt)\\
\xi_{y}  &  =\sqrt{2}\frac{\Delta S_{y}}{|\left\langle \mathbf{S}\right\rangle
|}\simeq1-\frac{4}{3}r\sin^{2}(\sqrt{\frac{3}{2}}gt)
\end{align*}
To the first order in $r$, the resulting state is a minimum uncertainty state,
and the minimum squeezing that can be achieved is the same for both
components. Squeezing as a function of $r$ is shown in Fig. \ref{Fig. 4}.%

\begin{figure}
[ptb]
\begin{center}
\label{Fig. 4}
\includegraphics[
trim=0.249398in 0.269322in 0.000000in 0.000000in,
natheight=3.032900in,
natwidth=3.884700in,
height=2.8046in,
width=3.6824in
]%
{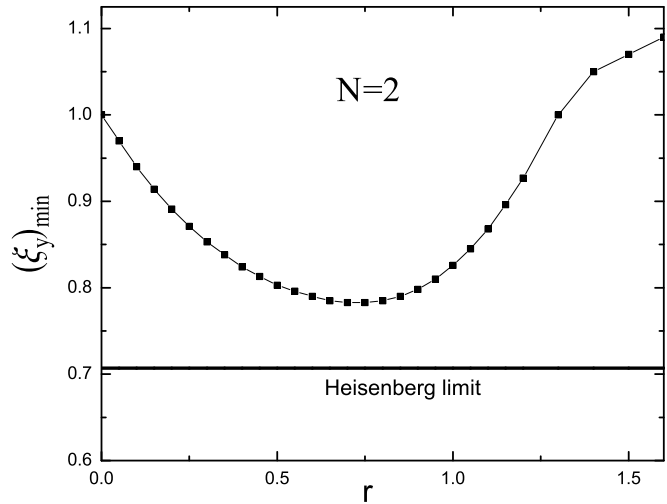}%
\caption{Optimal spin squeezing $\left(  \xi_{y}\right)  _{\min}$ as a
function of the squeezing parameter $r$ for an initially squeezed cavity field
for $N=2$. The time range out to $gt=5000$ was explored in obtaining the
minimal squeezing.}%
\end{center}
\end{figure}
With increasing $r$, $\xi_{y}$ decreases to minimum value of $0.78$ for
$r\approx0.7$, and then increases with increasing $r$. This result is
consistent with the general conclusion that optimal squeezing is obtained when
the average number of photons in the field is much less than $N.$

One might think that the squeezed vacuum produces optimal squeezing, but field
states that more closely approach the Heisenberg limit $\xi_{y}=1/\sqrt{2}$
can be constructed$.$ One such state is
\[
\left\vert \psi(0)\right\rangle =-0.79\left\vert 0\right\rangle
-0.594\left\vert 2\right\rangle +0.15\left\vert 4\right\rangle
+0.021\left\vert 6\right\rangle
\]
for which a minimum value $\xi_{y}=0.724$ is achieved. We have not been able
to formulate a general proof as to the minimum squeezing one can obtain for an
arbitrary initial state of the field.

\section{\textbf{\ N atoms}}

As the number of atoms, $N,$ increases, the spin squeezing that can be
achieved depends critically on the initial state of the cavity field. If the
field is in a coherent state, one might expect that the squeezing goes to zero
as $N$ goes to infinity since the atomic spin Hamiltonian approaches that of a
simple harmonic oscillator in this limit. A formal proof of this result is
given below. On the other hand, for finite $N,$ there are times for which spin
squeezing occurs, and the squeezing decreases with increasing field strength,
provided $N$ is much larger than the average number of photons in the field.
If the initial state of the field is a squeezed state such as the squeezed
vacuum, the field squeezing can be transferred to the atoms. In this manner,
one can generate a high degree of spin squeezing $\xi_{x}\ll1$, but still
considerably less than that predicted by the Heisenberg limit $\xi_{x}%
=1/\sqrt{N}$.

For arbitrary $N,$ the cavity field can, in principle, couple $\left(
N+1\right)  $ collective states corresponding to the angular momentum manifold
$S=N/2$. In practice, the number of states coupled is on the order of the
average number of photons in the initial field. The equations of motion for
the state amplitudes, obtained from Eqs. (\ref{wf}) and (\ref{hamil}) are%
\begin{equation}
\dot{c}_{mn}=-ig\left\{  \sqrt{\left(  \frac{N}{2}+m\right)  \left(  \frac
{N}{2}-m+1\right)  \left(  n+1\right)  }\,c_{m-1,n+1}+\sqrt{\left(  \frac
{N}{2}-m\right)  \left(  \frac{N}{2}+m+1\right)  n}\,c_{m+1,n-1}\right\}  ,
\label{recur}%
\end{equation}
with initial condition$\ c_{m,n}(0)=c_{n}\delta_{m,-N/2}$. This equation
represents a set of coupled equations, starting from $m=-N/2$ and reaching
some maximum value to $-N/2$ plus $n_{\max}$, where $n_{\max}$ is the smallest
$n$ where the initial field state amplitude $c_{n}$ is negligibly small.

\subsection{Coherent cavity field}

\subsubsection{\textbf{Analytical solution for }$\left\vert \alpha\right\vert
^{2}<<1$}

For $\alpha^{2}\ll1,$ the lowest order non-vanishing amplitudes obtained from
Eqs. (\ref{recur}) and (\ref{coh}) are%

\begin{subequations}
\label{cnl}%
\begin{align}
c_{-S,0}(t)  &  =(1-\alpha^{2}/2)\label{cnla}\\
c_{-S,1}(t)  &  =\alpha\cos(\sqrt{N}gt)\label{cnlb}\\
c_{-S+1,0}(t)  &  =-i\alpha\sin(\sqrt{N}gt)\ \label{cnlc}\\
c_{-S,2}(t)  &  =\frac{\alpha^{2}\sqrt{2}}{4N-2}[N-1+N\cos(\sqrt
{4N-2}gt)\ ]\label{cnld}\\
c_{-S+1,1}(t)  &  =-i\frac{\alpha^{2}\sqrt{N}}{\sqrt{4N-2}}\sin(\sqrt
{4N-2}gt)\label{cnle}\\
c_{-S+2,0}(t)  &  =-\frac{\alpha^{2}\sqrt{2N(N-1)}}{4N-2}[1-\cos(\sqrt
{4N-2}gt)\ ] \label{cnlf}%
\end{align}
In the large $N$ limit, the average spin components calculated using these
amplitudes are%

\end{subequations}
\begin{equation}
\left\langle S_{x}\right\rangle =0;\qquad\quad\left\langle S_{y}\right\rangle
=\sqrt{N}\alpha\sin(\sqrt{N}gt);\text{ \ \ }\left\langle S_{z}\right\rangle
=-\frac{N}{2}+\alpha^{2}\sin^{2}(\sqrt{N}gt)\ ,\text{\ \ } \label{spin3}%
\end{equation}
such that $\left\vert \left\langle \mathbf{S}\right\rangle \right\vert =S=N/2$
to order $\alpha^{2}.$

The squeezing parameter, calculated using Eqs. (\ref{sx}), is given by:%

\begin{equation}
\xi_{x}=\sqrt{N}\frac{\Delta S_{x}}{|\left\langle \mathbf{S}\right\rangle
|}\simeq1+\alpha^{2}\left\{  \frac{N-1}{N}\sin^{2}(\sqrt{N}gt)-\frac{2\left(
N-1\right)  }{2N-1}\sin^{2}\left(  \sqrt{(2N-1)/2}gt\right)  \right\}
\label{sxln1}%
\end{equation}
In the limit of large $N$ this reduces to%

\begin{equation}
\xi_{x}\sim1+\alpha^{2}\sin\left[  \left(  2\sqrt{N}-\frac{1}{4\sqrt{N}%
}\right)  gt\right]  \sin\left(  \frac{gt}{4\sqrt{N}}\right)  . \label{sxln2}%
\end{equation}
As $N$ approaches infinity, the squeezing vanishes; however, for any finite
$N$, there is a time of order $2\pi\sqrt{N}/g$ where spin squeezing with
$\xi_{x}\sim1-\alpha^{2}$ occurs. Note that, for small $gt\ll N^{-1/2},$
$\xi_{x}$ from (\ref{sxln1}) varies as $\left[  1-\alpha^{2}\left(  gt\right)
^{4}(N-1)/6\right]  $ while $\xi_{x}$ from (\ref{sxln2}) varies as $\left[
1+\alpha^{2}\left(  gt\right)  ^{2}/2\right]  $, which have different
functional forms; however, the \emph{difference }between these two results
varies as $\alpha^{2}/N\ll1/N\ll1$.

\subsubsection{\textbf{Numerical results for all values of }$\alpha$}

\begin{figure}
\label{Fig. 5}
[ptb]
\begin{center}
\includegraphics[
trim=0.249538in 0.159034in 0.000000in 0.000000in,
natheight=3.130600in,
natwidth=4.024800in,
height=3.0147in,
width=3.8242in
]%
{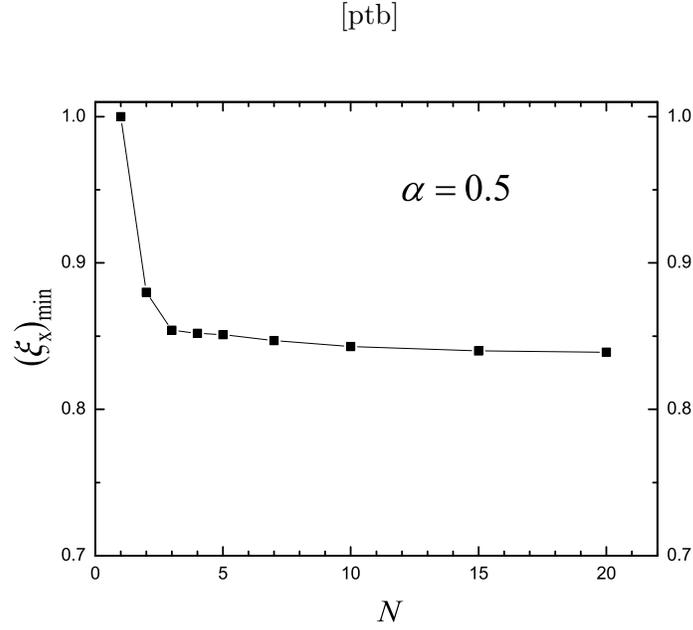}%
\caption{Optimal spin squeezing $\left(  \xi_{x}\right)  _{\min}$ as a
function of $N$ for $\alpha=0.5$.}%
\end{center}
\end{figure}

Since the average number of photons in a coherent state is $\alpha^{2},$ one
needs to solve Eq. (\ref{recur}) up to terms with $n\gg\alpha^{2}$. As
$\alpha$ grows the numerical solution becomes somewhat unwieldy. In Fig.
\ref{Fig. 5}, the optimal squeezing is plotted as a function of $N$ for
$\alpha=0.5.$%

\begin{figure}
\label{Fig. 6}
[ptb]
\begin{center}
\includegraphics[
trim=0.210082in 0.239886in 0.484269in 0.076192in,
natheight=5.952500in,
natwidth=7.723600in,
height=3.1055in,
width=3.87in
]%
{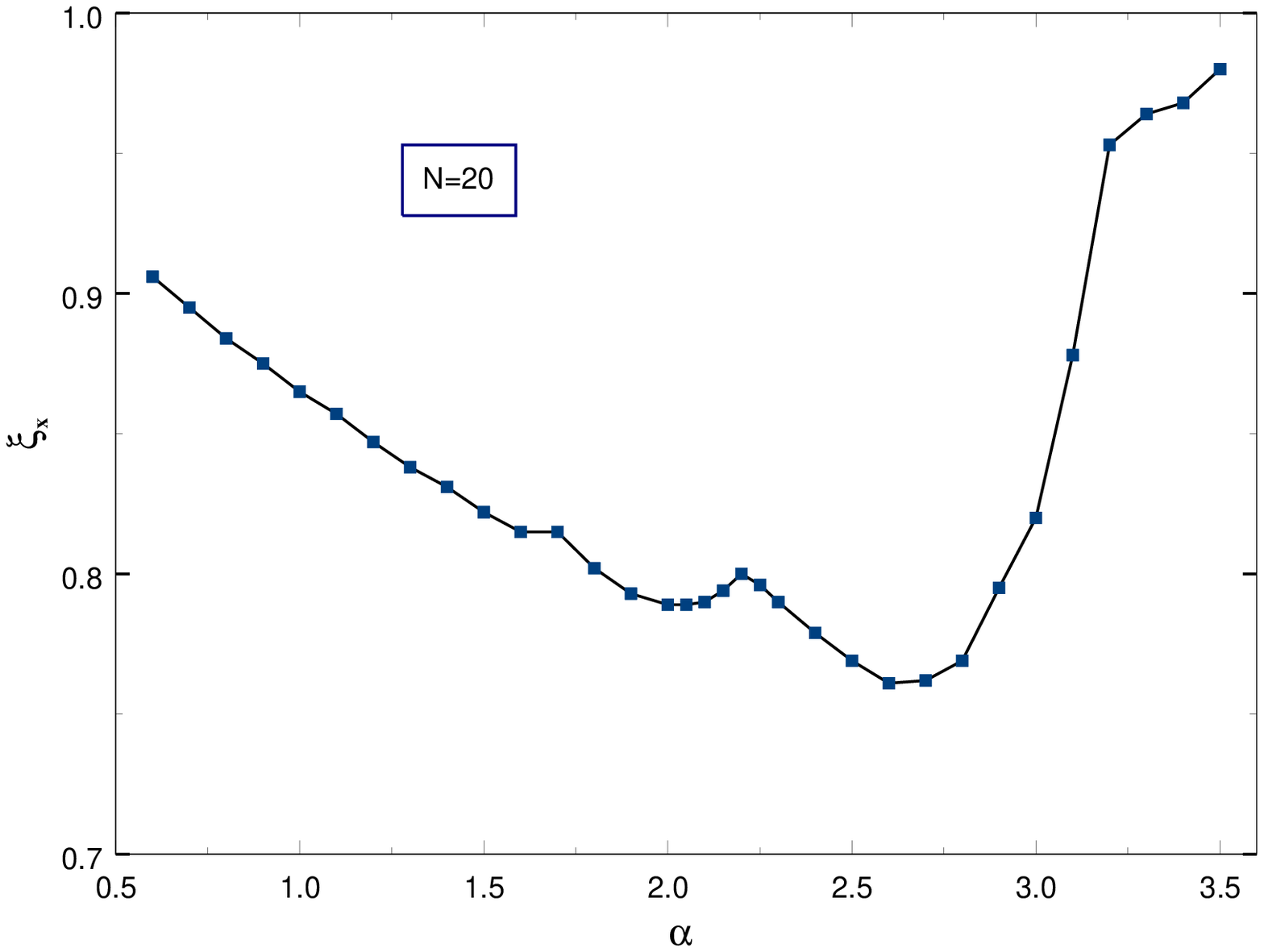}%
\caption{ Optimal spin squeezing $\left(  \xi_{x}\right)  _{\min}$ as a
function of $\alpha$ for $N=20$ and $0\leq gt\leq10$. Since only a restricted
range of $gt$ was considered, the values plotted may not represent the global
optimal squeezing, but still reflect the qualitative variation of $\left(
\xi_{x}\right)  _{\min}$ with $\alpha.$}%
\end{center}
\end{figure}

The squeezing diminishes with increasing $N$, eventually reaching an
asymptotic value of $0.86.$ This result represents the general trend that the
squeezing saturates for $N\gg\alpha^{2}.$ Spin squeezing as a function of
$\alpha$ for fixed $N=20$ is shown in Fig. \ref{Fig. 6}%

for $0.6\leq\alpha\leq3.5$. The values of $\xi_{x}$ in Fig. 6 are do not
necessary represent the optimal spin squeezing; rather they give first minimum
of the \emph{envelope} of a graph of $\xi_{x}$ versus $gt$. It is possible
that better spin squeezing occurs at higher values of $gt$ than those
considered (e.g., for $\alpha=0.6$, the first envelope minimum at $gt=9.03$
gives $\xi_{x}=0.906$, while the second envelope minimum at $gt=28.1$ gives
$\xi_{x}=0.817$); the computation time that would be needed to determine
$\left(  \xi_{x}\right)  _{\min}$ for all values of $gt$ grows rapidly with
increasing $\alpha$. Spin squeezing improves with increasing $\alpha$ up to
$\alpha\approx2.7\approx O(\sqrt{20})$ and then decreases with increasing
$\alpha$, following the general trend noted above.. Spin squeezing for larger
values of $\alpha$ and $N\gg\alpha^{2}$ are better treated by the method given
in Sec. V.

\subsubsection{\bigskip\textbf{Squeezing in the field}}

For $\alpha\ll1$, one finds squeezing parameters

\begin{align*}
\xi_{Q}  &  =\sqrt{2}\Delta\widehat{Q}\simeq1-\alpha^{2}\cos^{2}(\sqrt
{N}gt)+\frac{\alpha^{2}}{2N-1}\left[  N-1+N\cos(\sqrt{4N-2}gt)\right] \\
\xi_{P}  &  =\sqrt{2}\Delta\widehat{P}\simeq1+\alpha^{2}\cos^{2}(\sqrt
{N}gt)-\frac{\alpha^{2}}{2N-1}\left[  N-1+N\cos(\sqrt{4N-2}gt)\right]  ,
\end{align*}
implying that
\begin{subequations}
\label{spqzn}%
\begin{align}
(\xi_{Q})_{\min}  &  =1-\frac{2N}{2N-1}\alpha^{2}\label{spqzna}\\
(\xi_{P})_{\min}  &  =1-\alpha^{2}. \label{spqznb}%
\end{align}
The best squeezing is obtained for $N=2$. With increasing $\alpha$, the field
squeezing mirrors the spin squeezing.

\subsection{Squeezed initial cavity field}

The spin squeezing one can achieve increases dramatically if the initial state
of the cavity field is a squeezed state. For a squeezed vacuum with squeezing
parameter $r$, the initial squeezing in one quadrature component of the field
is $\xi_{Q}=e^{-r}.$ In the limit that $N\gg\sinh^{2}r+\sqrt{2}\sinh r\cosh
r=$(average plus standard deviation of the number of photons in the original
cavity field), one can show \cite{wineland} that this squeezing can be
transferred totally to the spins $\xi_{x}=e^{-r}$. For large $r$, this
represents substantial squeezing, but since $N\gg\sinh^{2}r+\sqrt{2}\sinh
r\cosh r,$ it follows that $\xi_{x}=e^{-r}$ $\gg\frac{1+\sqrt{2}}{2\sqrt{N}}$.
Thus, one is still far from the Heisenberg limit. It may be possible to
construct an original cavity field state that leads more closely to the
Heisenberg limit $\xi_{x}=1/\sqrt{N}$, but we have not explored this
possibility in the large $N\ $limit$.$

\section{\textbf{Asymptotic Solution for Large N}}

For an ensemble having a number of atoms much larger than unity and much
larger than the average number of photons in the field, the interaction
between the atoms and the cavity field can be seen as an interaction between a
harmonic oscillator (the field) and an imperfect oscillator (the atoms). To
attempt to map this problem into one of interacting harmonic oscillators,
which will be valid as the number of atoms $N$ approaches infinity, one
defines boson operators for the atoms via
\end{subequations}
\begin{subequations}
\label{hp}%
\begin{align}
S_{z}  &  =-N/2+b^{\dagger}b,\label{hpa}\\
S^{+}  &  =e^{-i\omega t}{N}^{1/2}b^{\dagger}(1-b^{\dagger}b/N)^{1/2}\simeq
e^{-i\omega t}\left(  \sqrt{N}b^{\dagger}-\frac{1}{2\sqrt{N}}b^{\dagger
}b^{\dagger}b\right)  ,\label{hpb}\\
S^{-}  &  =e^{i\omega t}{N}^{1/2}(1-b^{\dagger}b/N)^{1/2}b\simeq e^{i\omega
t}\left(  \sqrt{N}b-\frac{1}{2\sqrt{N}}b^{\dagger}bb\right)  . \label{hpc}%
\end{align}
The boson occupation states (Fock states) $|m\rangle=\frac{(b^{\dagger})^{m}%
}{\sqrt{m!}}|0\rangle$ correspond to the different projections onto the
collective angular momentum states and, in effect, represent excitations above
the lowest state having $S_{z}=-N/2.$ The transformation to the $b$ bosons
(Holstein-Primakoff transformation \cite{holstein}) is exact. The
approximations in (\ref{hpa}) and (\ref{hpb}) are valid provided the relative
variations of the spin projection are small:
\end{subequations}
\begin{equation}
\langle b^{\dagger}b\rangle/N\ll1; \label{li}%
\end{equation}
in other words, the average spin remains aligned very close to the $z$ axis.
The key point in this calculation is that all changes in the eigenkets of
order $1/\sqrt{N}$ are neglected. Changes in the eigenenergies of order
$1/\sqrt{N}$ lead to significant changes in the \emph{phases} of the
time-dependent wave function for any finite $N$. Such changes in the phase can
result in spin squeezing.

The total Hamiltonian ( in an interaction representation) is written as
$H=H_{0}+H^{\prime}$, with
\begin{subequations}
\label{hr}%
\begin{align}
H_{0}  &  =\hbar\sqrt{N}g\left(  b^{\dagger}a+a^{\dagger}b\right)
,\label{hra}\\
H^{\prime}  &  =-\frac{\hbar g}{2\sqrt{N}}\left(  b^{\dagger}b^{\dagger
}ba+a^{\dagger}b^{\dagger}bb\right)  . \label{hrb}%
\end{align}
We now diagonalize $H_{0}$ and treat $H^{\prime}$ as a perturbation. The
Hamiltonian $H_{0}$ can be written as%

\end{subequations}
\[
H_{0}=\omega_{+}\Gamma^{\dagger}\Gamma+\omega_{-}\gamma^{\dagger}%
\gamma;\;\;\;\;\;\;\;\;\;\;\;\omega_{\pm}=\pm\sqrt{N}g,
\]
with%
\begin{align*}
\Gamma^{\dagger}  &  =\frac{a^{\dagger}+b^{\dagger}}{\sqrt{2}};\text{
\ \ \ \ }\gamma^{\dagger}=\frac{a^{\dagger}-b^{\dagger}}{\sqrt{2}},\\
a^{\dagger}  &  =\frac{\Gamma^{\dagger}+\gamma^{\dagger}}{\sqrt{2}};\text{
\ \ \ \ }b^{\dagger}=\frac{\Gamma^{\dagger}-\gamma^{\dagger}}{\sqrt{2}},
\end{align*}
while perturbation $H^{\prime}$ has the form
\[
H^{\prime}=\frac{\hbar g}{4\sqrt{N}}\left[  \gamma^{\dagger}\gamma^{\dagger
}\gamma\gamma-\Gamma^{\dagger}\Gamma^{\dagger}\Gamma\Gamma+\Gamma^{\dagger
}\Gamma^{\dagger}\gamma\gamma+\gamma^{\dagger}\gamma^{\dagger}\Gamma
\Gamma+2\left\{  \Gamma^{\dagger}\gamma^{\dagger}(\gamma\gamma-\Gamma
\Gamma)+h.c.\right\}  \right]  .
\]
Only the first two terms in this expression, conserving the total number of
excitations, contribute in first order.

The eigenkets of $H_{0}$ are
\begin{equation}
|n\rangle_{+}|m\rangle_{-}=\frac{(\Gamma^{\dagger})^{n}}{\sqrt{n!}}%
|0\rangle\frac{(\gamma^{\dagger})^{m}}{\sqrt{m!}}|0\rangle, \label{h0e}%
\end{equation}
with energies
\[
\epsilon^{(0)}(n,m)=\hbar\left(  \omega_{+}n+\omega_{-}m\right)  .
\]
The first order correction to the energies of these states is
\[
\epsilon^{(1)}(n,m)=\frac{\hbar g}{4\sqrt{N}}(n-n^{2}-m+m^{2})\equiv
\epsilon_{+}^{(1)}(n)+\epsilon_{-}^{(1)}(m),
\]
and we define
\[
\epsilon_{\pm}(n)=\hbar\omega_{\pm}n+\epsilon_{\pm}^{(1)}(n).
\]
To this order the states (\ref{h0e}) are unmodified.

In order to neglect higher order correction to the energies, it is necessary
that the phase produced by such corrections must be much less than unity. This
translates into the condition $\frac{g^{2}}{N\left(  \omega_{+}-\omega
_{-}\right)  }t=\frac{gt}{2N^{3/2}}\ll1$, which can always be satisfied for
sufficiently large $N,$ but would be violated for $N=2.$ There is no
restriction on the value of the phase $\left\vert \epsilon_{\pm}%
(n+1)-\epsilon_{\pm}(n)\right\vert t/\hbar\approx ngt/\sqrt{N}$, provided
$gt/\left(  2N^{3/2}\right)  \ll1.$ In fact, such phases are responsible for
the finite $N$ corrections calculated below.

\subsection{Coherent cavity field}

For an initial state in which the cavity field is in a coherent state and
$S_{z}=-N/2$, one has
\[
|\Psi(t=0)\rangle=e^{-\tilde{\alpha}^{2}}e^{\sqrt{2}\tilde{\alpha}a^{\dagger}%
}|0\rangle=e^{-\tilde{\alpha}^{2}}e^{\tilde{\alpha}(\Gamma^{\dagger}%
+\gamma^{\dagger})}|0\rangle\ ,
\]

\[
|\Psi(t)\rangle=e^{-\tilde{\alpha}^{2}}\left(  \sum_{m}\frac{\tilde{\alpha
}^{m}}{\sqrt{m!}}e^{-i\epsilon_{+}(m)t/\hbar}e^{-im\omega t}|m\rangle
_{+}\right)  \left(  \sum_{n}\frac{\tilde{\alpha}^{n}}{\sqrt{n!}}%
e^{-i\epsilon_{-}(n)t/\hbar}e^{-in\omega t}|n\rangle_{-}\right)  ,
\]

\begin{align*}
\langle\Psi(t)|b|\Psi(t)\rangle &  =\frac{e^{-\tilde{\alpha}^{2}}}{\sqrt{2}%
}\left\{  \sum_{m}\langle m-1|\Gamma|m\rangle_{_{+}}\frac{\tilde{\alpha}%
^{m-1}\tilde{\alpha}^{m}}{\sqrt{m!(m-1)!}}e^{-i\omega t}e^{-i[\epsilon
_{+}(m)-\epsilon_{+}(m-1)]t/\hbar}\right. \\
&  \left.  -\sum_{n}\langle n-1|\gamma|n\rangle_{_{-}}\frac{\tilde{\alpha
}^{n-1}\tilde{\alpha}^{n}}{\sqrt{n!(n-1)!}}e^{-i\omega t}e^{-i[\epsilon
_{-}(n)-\epsilon_{-}(n-1)]t/\hbar}\right\} \\
&  =\tilde{\alpha}e^{-i\omega t}{e^{-\tilde{\alpha}^{2}}}{\sqrt{2}i}\sum
_{n}\frac{\tilde{\alpha}^{2(n-1)}}{(n-1)!}\sin\left[  (\lambda_{1}%
-n\lambda_{2})\right]  t,
\end{align*}
with $\tilde{\alpha}=\alpha/\sqrt{2}$ and
\[
\lambda_{1}=\left(  \sqrt{N}+\frac{1}{2\sqrt{N}}\right)
g,\;\;\;\;\;\;\;\lambda_{2}=\frac{1}{2\sqrt{N}}g.
\]
Note that%

\[
\left\langle S_{x}\right\rangle =\sqrt{N}\left[  \langle be^{i\omega t}%
\rangle+\langle b^{\dagger}e^{-i\omega t}\rangle\right]  =0.
\]

In order to compute the squeezing, we need the following averages:
\begin{align*}
\langle\Gamma^{\dagger}\Gamma^{\dagger}\rangle &  =\tilde{\alpha}%
^{2}e^{i2\omega t}{e^{-\tilde{\alpha}^{2}}}e^{i(2\lambda_{1}-3\lambda_{2}%
)t}\sum_{n}\frac{\left(  \tilde{\alpha}^{2}e^{-2i\lambda_{2}t}\right)  ^{n}%
}{n!}\\
&  =\tilde{\alpha}^{2}e^{i2\omega t}e^{i(2\lambda_{1}-3\lambda_{2})t}%
e^{\tilde{\alpha}^{2}(e^{-2i\lambda_{2}t}-1)};\\
\langle\gamma^{\dagger}\gamma^{\dagger}\rangle &  =\tilde{\alpha}%
^{2}e^{i2\omega t}e^{-i(2\lambda_{1}-3\lambda_{2})t}e^{\tilde{\alpha}%
^{2}(e^{2i\lambda_{2}t}-1)};\\
\langle\Gamma^{\dagger}\rangle &  =\tilde{\alpha}e^{i\omega t}e^{i\sqrt{N}%
gt}e^{\tilde{\alpha}^{2}(e^{-i\lambda_{2}t}-1)};\\
\langle\gamma^{\dagger}\rangle &  =\tilde{\alpha}e^{i\omega t}e^{-i\sqrt{N}%
gt}e^{\tilde{\alpha}^{2}(e^{i\lambda_{2}t}-1)}.
\end{align*}
The value of $\left\langle \mathbf{S}\right\rangle $ remains equal to $N/2$,
with corrections of order $\alpha^{2}/N$, and the squeezing, $\xi_{x}%
\approx(2/\sqrt{N})\Delta S_{x}$ is calculated as
\begin{align}
\xi_{x}  &  =\sqrt{\langle\left(  {b^{\dagger}e^{-i\omega t}+be^{i\omega t}%
}\right)  ^{2}\rangle}\nonumber\\
&  =\left\{  1+\alpha^{2}\left[  e^{-\alpha^{2}\sin^{2}\left(  \lambda_{{2}%
}t\right)  }\cos\left(  (2\sqrt{N}g-\lambda_{2})t-\left(  \alpha^{2}/2\right)
\sin\left(  2\lambda_{2}t\right)  \right)  -e^{-2\alpha^{2}\sin^{2}\left(
\lambda_{2}t/2\right)  }\right.  \right. \nonumber\\
&  +\left.  \left.  1-e^{-2\alpha^{2}\sin^{2}\left(  \lambda_{2}t/2\right)
}\cos\left(  2\sqrt{N}gt-\alpha^{2}\sin\left(  \lambda_{2}t\right)  \right)
\right]  \right\}  ^{1/2}. \label{sxr}%
\end{align}
This expression agrees with Eq. (\ref{sxln2}) in the limit that $\alpha\ll1;$
however, it extends that result to all values of $\alpha$ for which condition
(\ref{li}) remains valid and for which $gt/\left(  2N^{3/2}\right)  \ll1$
\cite{small}$.$ Persico and Vetri \cite{persico} employ a somewhat different
approach in solving this problem using the Holstein-Primakoff transformation
and obtain a validity range, $gt<\sqrt{N}/\alpha^{2}.$ Since $2N^{3/2}\gg
\sqrt{N}/\alpha^{2}$, the validity range for Eq. (\ref{sxr}) should be much
greater than that of Persico and Vetri. To test this hypothesis, we compared
the term of order $\alpha^{4}$ in the exact solution with the $\alpha^{4}$
term of (\ref{sxr}). The two results agreed for times $gt/\left(
2N^{3/2}\right)  \ll1,$ as expected. It might be noted that Eq. (\ref{sxr}),
agrees with the exact result to order $\alpha^{2}$, independent of $gt$,
provided $N\gg1.$ This is why we had to compare the $\alpha^{4}$ terms.

\begin{figure}
[ptb]
\begin{center}
\label{Fig. 7}
\includegraphics[
height=2.2044in,
width=3.5648in
]%
{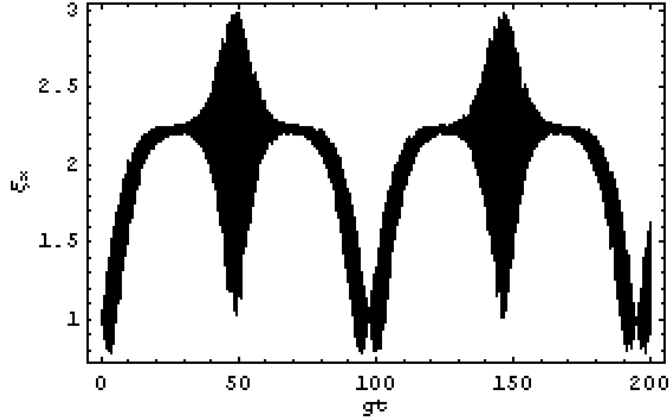}%
\caption{Spin squeezing $\xi_{x}$ as a function of $gt$ for $\alpha=2$ and
$N=60$.}%
\end{center}
\end{figure}

For $\alpha\ll1$, there is a slow modulation having period\ $gt=4\pi\sqrt{N}$,
in addition to the rapid oscillations having period $gt=\pi/\sqrt{N}.$ With
increasing $\alpha,$ and $N\gg\alpha^{2},$ the overall period is $gt=4\pi
\sqrt{N},$ with a subharmonic having period $gt=2\pi\sqrt{N}$, and the rapid
oscillations having period $gt=\pi/\sqrt{N}.$ These features are seen clearly
in Fig. \ref{Fig. 7}, drawn for $\alpha=2$ and $N=60.$%

\begin{figure}
[ptb]
\begin{center}
\label{Fig. 8}
\includegraphics[
natheight=3.102100in,
natwidth=3.982500in,
height=3.1479in,
width=4.0318in
]%
{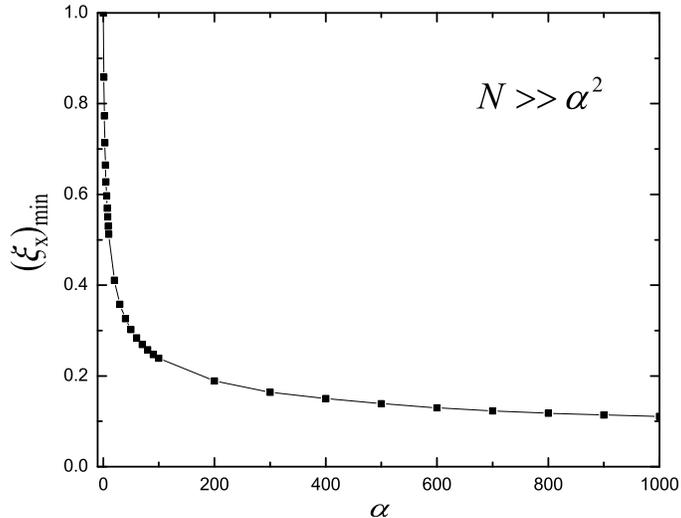}%
\caption{Optimal spin squeezing $\left(  \xi_{x}\right)  _{\min}$ as a
function of $\alpha$ for $N\gg\alpha^{2}$.}%
\end{center}
\end{figure}

Similar curves were obtained by Kozierowski and Chumakov \cite{prevtc} for the
field squeezing. With increasing $\alpha,$ the maximum squeezing increases
slowly as is shown in Fig. \ref{Fig. 8}, where the condition $N\gg\alpha^{2}$
is maintained as $\alpha$ is varied.%

In contrast to the $\alpha\ll1$ case, the optimal squeezing for $\alpha\gg1$,
always occurs at a time $gt\approx\sqrt{N}/\alpha^{3/2}\ll\sqrt{N}.$ In the
limit that $\alpha\gg1$ and $z\equiv\alpha^{2}gt/2\sqrt{N}\ll\sqrt{\alpha}$,
one can show that Eq. (\ref{sxr}) can be approximated as
\[
\xi_{x}\approx\left\{  1+z\sin\left(  \sigma z-z\right)  +z^{2}\sin^{2}\left[
\left(  \sigma z-z\right)  /2\right]  \right\}  ^{1/2},
\]
where $\sigma=4N/\alpha^{2}.$ From this expression it is possible to show that
the squeezing parameter goes to zero with increasing $\alpha$, but that the
approach to zero is slower than $\alpha^{-1/2}$ (the actual dependence seems
to be close to $\alpha^{-0.31}).$ Even though the field is getting more
classical with increasing $\alpha$, quantum fluctuations in the field still
lead to increased squeezing with increasing $\alpha.$ Of course, if we explore
the range $\alpha^{2}>N,$ we would find a decrease in squeezing with
increasing $\alpha,$ as we found for the case $N=2$.

\section{Summary}

It has been shown that a linear interaction Hamiltonian between a coherent
state cavity field and an ensemble of two-level atoms can produce spin
squeezing. Analytical solutions for small values of the amplitude of the field
state were derived, showing a reduction is the squeezing parameter quadratic
in $\alpha$. Computer simulations were used to find the best value for
squeezing, when $\alpha$ is varied over a range of real, positive values. The
limit of a large number of atoms was also examined. For an initial coherent
state for the cavity field, it was found that the squeezing approaches zero
with increasing $\alpha$. This might seem like a remarkable result since the
coherent state closely resembles a classical field for large $\alpha$. Even
though $\alpha$ is large, the number of atoms is assumed to be much larger
than $\alpha^{2}$; as such the field can be totally depleted. The entanglement
of the field and the spins can produce significant phase shifts that can lead
to spin squeezing. Although $\xi_{x}$ approaches zero with increasing $\alpha
$, the ratio $\xi_{r}=\xi_{x}/\sqrt{N}$ that relates the squeezing to the
Heisenberg limit, \emph{decreases} with increasing $\alpha$. If squeezing
relative to the Heisenberg limit is used as a measure, the best squeezing is
obtained for $N=2.$ This is in marked contrast to the optimal squeezing that
can be obtained with nonlinear spin interactions \cite{wineland,theory}.

The interaction with a squeezed cavity field was also investigated. While a
squeezed vacuum field has the potential to transfer significant spin squeezing
to the atoms, the degree of spin squeezing produced is still well above the
Heisenberg limit. By constructing alternative squeezed states, we were able to
improve the squeezing relative to that of a spin-squeezed vacuum, but the
ultimate degree of spin squeezing that can be transferred to the atoms via an
interaction with a cavity field remains an open question.

\bigskip

\section{Acknowledgments}

This work is supported by the U. S. Office of Army Research under Grant No.
DAAD19-00-1-0412 and by the National Science Foundation under Grant No.
PHY-0098016 and the FOCUS Center Grant.

\bigskip

\bigskip\ 

\newpage

Figure Captions

\bigskip

Fig. 1. Spin squeezing $\xi_{x}$ as a function of $gt$ for $\alpha=0.4$ and
$N=2$.

Fig. 2. Optimal spin squeezing $\left(  \xi_{x}\right)  _{\min}$ as a function
of $\alpha$ for $N=2$. The time range out to $gt=5000$ was explored in
obtaining the minimal squeezing. In this and other plots, the point represent
actual values for which the squeezing was calculated. A line is drawn through
these points.

Fig. 3. Optimal field squeezing $\left(  \xi_{Q}\right)  _{\min}$ as a
function of $\alpha$ for $N=2$. The time range out to $gt=5000$ was explored
in obtaining the minimal squeezing.

Fig. 4. Optimal spin squeezing $\left(  \xi_{y}\right)  _{\min}$ as a function
of the squeezing parameter $r$ for an initially squeezed cavity field for
$N=2$. The time range out to $gt=5000$ was explored in obtaining the minimal squeezing.

Fig. 5. Optimal spin squeezing $\left(  \xi_{x}\right)  _{\min}$ as a function
of $N$ for $\alpha=0.5$.

Fig. 6. Optimal spin squeezing $\left(  \xi_{x}\right)  _{\min}$ as a function
of $\alpha$ for $N=20$ and $0\leq gt\leq10$. Since only a restricted range of
$gt$ was considered, the values plotted may not represent the global optimal
squeezing, but still reflect the qualitative variation of $\left(  \xi
_{x}\right)  _{\min}$ with $\alpha.$

Fig 7. Spin squeezing $\xi_{x}$ as a function of $gt$ for $\alpha=2$ and
$N=60$.

Fig. 8. Optimal spin squeezing $\left(  \xi_{x}\right)  _{\min}$ as a function
of $\alpha$ for $N\gg\alpha^{2}$.


\begin{thebibliography}{9}
\bibitem {small}Note that Eq. (\ref{sxr}) differs from Eq. (\ref{sxln1}) if
$gt\ll N^{-1/2}$ by terms of order $\alpha^{2}/N\ll1/N$; terms of order $1/N$
neglected in arriving at Eq. (\ref{sxr}) can contribute in absolute terms if
$\left\vert \xi_{x}-1\right\vert \ll1.$
\end{thebibliography}
\end{document}